\documentstyle[12pt]{article}

\newcommand{\noind}{\noindent}
\newcommand{\be}{\begin{equation}}
\newcommand{\ee}{\end{equation}}

\begin{document}
 \thispagestyle{empty}

\noind{\bf July,  1992} \hfill{UM-P-92/61}

\hfill{OZ-92/20}
\begin{center}

{\Large \bf   Quaternions and the Heuristic Role of \\
Mathematical Structures in Physics }

 \vspace{7mm}

 {\bf Ronald Anderson, S.J.}

 {\it Department of Philosophy, Boston College}

 {\it Chestnut Hill MA 02167, U.S.A.}\\

\vspace{4mm}

 {\bf Girish C. Joshi}

{\it Centre for High Energy Physics, School of Physics}

{\it University of Melbourne, Parkville, Victoria 3052}

{\it Australia}\\

\vspace{6mm}
{\large\bf Abstract}
\end{center}
 \vspace{5mm}

\noind  One of the important ways development takes place in mathematics
is via a process of generalization.   On the basis of a recent characterization
of
this process we propose a principle that generalizations of mathematical
structures that are already part of successful physical theories serve as good
guides for the development of new physical theories.  The principle is a more
formal presentation and extension of a position stated  earlier this century by
Dirac.   Quaternions form an excellent example of such a generalization, and
we consider a number of the ways in which their use in physical theories
illustrates this
principle.

\vspace{5mm}

\noindent{\bf Key words}:  Quaternions, heuristics, mathematics and
physics,
quaternionic quantum theory.

\newpage
\section{Introduction}

In recent decades the necessary role mathematical structures play in the
formulation of physical theories has been the subject of ongoing interest.
Wigner's reference in a well known essay of 1960~\cite{Wigner;1960} to the
``unreasonable effectiveness'' of mathematics in this role has captured what
is undoubtedly a widespread feeling that this success is remarkable, and
moreover, in need of further explanation.  Wigner himself noted that this
role of mathematics is a ``wonderful gift we neither understand nor
deserve.''  The topic has been of interest not only to physicists and
mathematicians, but also to those working on the philosophical implications
and foundations of both subjects.\footnote{
The recent collection of essays in Ref.~\cite{Mickens;1990} provides a
guide to the literature as well as an introduction to the variety of
ways in which this topic may be approached.   Other recent discussions
may be found in
Refs.~\cite{Barrow;1988,Davies;1992,Hellman;1989,Penrose;1989}.}

	Much of the discussion on this topic has focused on particular physical
theories and sought to explore what one may infer about the nature of either
mathematical or physical knowledge or  the entities of concern to both
disciplines from the role of mathematics in these theories.  Mathematics,
however, also plays an important role in the {\em development} of new
physical theories, and while less attention has been paid to this heuristic
role
of mathematics, its importance has been well recognized.   In a series of
essays
Bochner~\cite{Bochner;1966}, for example, has traced significant episodes in
the
development of physics where mathematics has played a crucial role, and in
recent studies Redhead~\cite{Redhead;1975} and
Zahar~\cite{Zahar;1980,Zahar;1989} have identified in a formal manner a
number of
ways in which this may take place.  In this essay we wish to propose a way in
which
mathematics may play such a heuristic role in physics which is not explicitly
mentioned in these works, although an implicit
recognition of it may be found in the work of Bochner~\cite{Bochner;1966}.
In particular we with to draw attention to the importance of certain
developments within {\em mathematics itself\/}  for the development of
physics.

Unlike the situation in the natural sciences, a concern with the particular
manner in which mathematics evolves has  been of relatively recent origin.
The works of Crowe~\cite{Crowe;1975}, Koppelman~\cite{Koppelman;1975}
and Wilder~\cite{Wilder;1968,Wilder;1981}, for example, which aim at
characterizing the nature of mathematical evolution, only stem from the late
sixties.  Of even more recent origin are the  works on this topic by
Hallet~\cite{Hallett;1979}, Kitcher~\cite{Kitcher;1984}, and
others~\cite{Glas;1989,Yuxin;1990} within the philosophy of mathematics.  In
these works the ideas of philosophers such as
Kuhn and Lakatos have been used
to explore  both the question of progress and the
``logic of discovery'' in mathematics as well as the parallels between the
development of mathematics and science.

In this essay we wish to exploit a characterization given by
Kitcher~\cite{Kitcher;1984} of one of the important ways mathematics
progresses, which he identifies as one of ``generalization.'' This refines an
idea
stated by Dirac in 1931 on the manner in which certain developments in
mathematics can play an important heuristic role in physics.   In particular,
we wish to propose that generalizations of those mathematical structures of
physical theories which at any stage enjoy a measure of success in describing
nature supply new mathematical structures that can serve vital roles in the
development of new physical  theories.  In addition, we propose that through
their use in this manner better  understanding of present theories is
obtained which in turn can give a good base for the development of new
theories.  In Section 2 we will provide a
characterization of generalization which will flesh out this position, and we
will provide some examples of where it has occurred in the development of
physics.   Dirac's statement of the idea occurs in his famous essay of 1931 on
the
quantized singularities in the electromagnetic
field:\footnote{
The idea can also be found in general articles by Dirac
in Refs.~\cite{Dirac;1939,Dirac;1963}.   In Ref.~\cite{Dirac;1939} Dirac
notes that a ``powerful new method'' for the physicist consists of
choosing a branch of mathematics and then proceeding ``to develop it
along suitable lines, at the same time looking for that way in which it
appears to lend itself naturally to physical interpretation.''  While
mention is made in Ref.~\cite{Dirac;1963} of the need for a ``higher
and higher'' mathematics no mention is directly made of
``generalization'' in either article.  Instead ``mathematical beauty'' is
mentioned as one of the criteria for deciding on appropriate
mathematical developments.   Beauty in mathematics, however, is
notoriously difficult to define, while the notion of ``generalization'' is
more amenable to specification.}
\begin{quote}
The most powerful method of advance that can be suggested at present is to
employ all the resources of pure mathematics in attempts to perfect and
generalize the mathematical formalism that forms the existing basis of
theoretical physics, and {\em after\/} each success in this direction, to try
to
interpret the new mathematical features in terms of physical
entities\ldots \cite{Dirac;1931}
\end{quote}
Naturally, the task of interpreting the new mathematical structures and
forming a physical theory using the structures is an all important and
difficult
one, however, our focus in this essay is on the importance of exploiting
certain types of mathematical developments.   In addition, the statement of
our position in the next section must of necessity be somewhat informal.
Non-trivial generalizations in mathematics require creative insights that by
their
nature defy prediction, and moreover, there is no guarantee that all possible
generalizations of the structures in use at any one time in physics might be of
relevance to new physical theories.  Our investigation of the natural world is
such that we have no well-defined algorithm for generating new scientific
theories.   Nevertheless, there is good historical evidence that  mathematical
structures of the sort indicated above are indeed productive in the
development of new physical theories.

We should mention that others have expressed ideas similar to the one of
Dirac we have mentioned above.  Whitehead~\cite{Whitehead;1926}, for
example, writing on the role of mathematics in science around the same time
as Dirac noted a similar phenomenon in the manner in which the growth of
modern physics depended very much on advances towards abstraction in
mathematics:
\begin{quote}
Nothing is more impressive than the fact that as mathematics withdrew
increasingly into the upper regions of very greater extremes of abstract
thought, it returned back to earth with a corresponding growth of importance
for the analysis of concrete fact. \ldots The paradox is now fully established
that the utmost abstractions are the true weapons with which to control our
thought of concrete fact.\footnote{Ref.~\cite{Whitehead;1926}, p.\ 47.}
\end{quote}
\noindent The concept of generalization which we will outline in Section 2
is not
unlike the process of abstraction Whitehead is referring to, and moreover,
our proposal maintains in a similar manner to Whitehead the idea that
mathematical developments of this nature are the ones  best able to capture
the particularity of our world.

	In Sections 3--4 of our essay this position will be explored for the
particular ``test case'' example of quaternions.  Quaternions were discovered
by Hamilton in 1843.  They form an associative division algebra of which the
only other members are the real and complex numbers.  They provide an
excellent example of the type of mathematical generalizations of  concern to
us, and furthermore, they are generalizations of  complex and real numbers,
numbers which play central roles in current physical theories.  The history of
the attempts to use them in physical theories is both interesting and marked
with controversy, and as  G\"ursey~\cite{Gursey;1983}  has observed,  forms
an interesting episode in the relationship between mathematics and physics.

We will argue that while quaternions at the moment do not have an
assured a place in physical theories as do the other two associative division
algebras, there are a number of  interesting senses in which they do illustrate
our principle.    Our treatment of quaternions is not intended to be
comprehensive as the history of their use in physics has been well covered in
other places~\cite{Gursey;1983}; rather, we will focus directly on aspects
related to our argument concerning the role of mathematical structures in
physics.

 \section{ Mathematical Generalization and the
Development  of Physics}

\subsection{Mathematical Generalization}

The patterns evident in mathematical evolution are numerous and complex.
Kitcher \cite{Kitcher;1984} identifies five types as being of importance for
the
progress of mathematics one of which is the process of generalization.
For Kitcher this process entails several elements: i) it introduces new
expressions to
the
mathematical language;  ii) it preserves some features used in the old
expressions;  iii) in the process certain constraints on prior usage are
abandoned; iv)  a new theory is obtained  with analogues of the old; v) the
new structure brings to our attention properties of  familiar entities which
enables us to see the old theory as a special case.  In this way the
generalization enables areas already developed to be illuminated in a new
way.  Kitcher mentions the examples of Hamilton's creation of quaternions,
Lebesgue's theory of integration, and Cantor's extension of arithmetic to
transfinite numbers as fitting his characterization of generalization.

While Kitcher focuses on characterizing generalizations as evident in the
work of given individuals, there is good reason to see many developments in
mathematics as fitting this pattern, even though neither at the point of their
creation nor in their development by their creator might all of these elements
have been evident.  For example, it was not until the middle of the
nineteenth century,
after they had established themselves in algebra and in the theory of complex
functions through the geometrical interpretation of Wessel, Argand and
Gauss, that complex numbers could be seen as  generalizations of real
numbers in
the sense given in
 i)--v).\footnote{Details of this development may be found in Kline's
excellent history of mathematics~\cite{Kline;1972}.}
  The development of non-Euclidean geometries
can also be seen as a generalization of this sort as well as the developments
that arose from the move of ``localization'' made by Riemann in the study of
the
geometry of surfaces, and Lie in the study of continuous groups.

With this specification of generalization of a mathematical structure, Dirac's
idea is made more precise.    The reasons as to why generalizations of this
nature of structures which are currently part of successful physical theories
should be of
such
importance for the development of new physics is an important question, but
it is not one we address here.  We feel there is sufficient evidence from the
actual development of physics to justify such a position, and  that a
profitable
way to go about developing new physical theories will be to proceed in the
manner the principle suggests. We choose two particular mathematical
structures as illustrations.

\subsection{Some Illustrations}

Our first is pertinent to the theme of the paper and concerns two roles
complex numbers have had in the development of physics through their
ability to represent phases.  Here we view complex numbers as
generalizations of real numbers.    In a work published in 1831 Fresnel noted
that for
certain angles of incidence and reflection of polarized light the ratio of the
amplitudes is complex but with an absolute value of 1.  Fresnel interpreted
the
ratio to be given by  $e^{i\theta}$ with $\theta$ representing the phase shift
between the two waves.  His result was later experimentally confirmed.
Another place where complex numbers play an essential role is in quantum
mechanics, and moreover their presence uniquely characterizes important
features of the theory.  At the time of the original formulations of both
Heisenberg and Schr\"odinger, their presence posed problems of how to
interpret the mathematical formalism of the theory.  These were solved by
Born's
interpretation of taking $\psi\psi^{*}$ to represent a probability density.
In its most common modern formulation states are represented in a complex
vector space, and via their presence as complex phases represent  interference
phenomena that are unique to quantum states.   In addition, as it was
recognized very
early in the development of quantum theory, electromagnetism could be
incorporated into the theory via the complex phases.
Thus the feature noted by Argand in 1906 that complex numbers generalize
real numbers by adding a concept of rotation has proved to be of extraordinary
value in providing an added structure needed for theories such as quantum
mechanics.\footnote{
For Bochner~\cite{Bochner;1966}, Fresnel's achievement was the first
time physical features were ``abstracted'' from a purely mathematical
structure that had been developed independently of physical
considerations.  Bochner also clearly notes that complex numbers
provide a higher level of abstraction than real numbers.   In
combination both of these positions can be seen as implying the one we
are maintaining here.}

Our second example concerns the important role Lie groups play in
physical theories, and we note two places where generalizations within the
concept of Lie groups have been important in developing new physical
theories.   The first was the use of a non-Abelian Lie group by Yang and Mills
in 1954 to form a theory built on a generalization of  electromagnetic gauge
invariance. Thus instead of a single scalar function characterizing the
transformation, as in the case of electromagnetism,  the functions were
members of a
non-Abelian group.  The work of Yang and Mills   has flowered into the
current gauge theories of the fundamental interactions.  A second place
where a generalization has been important has been in the development of
supersymmetry theories.  These are based on a generalization of the notion of
a
Lie algebra to a graded Lie algebra~\cite{West;1990}.  While there is no
evidence that such a symmetry is realized in nature, supersymmetry theories
have many attractive features such as relating spacetime and internal
symmetries in a non-trivial manner and providing a Fermi-Bose symmetry.

We are not arguing that these theories proceeded by the deliberate use of
mathematics in the sense we are proposing; the complex development
evident in all of the above examples reveals a variety of motives and ways of
proceeding.  It does, however, reveal a pattern in the type of  mathematics
that can be important in the development of new physical theories.   For our
principle to have a ``force'' to it, it is important that all the elements of
generalization of Kitcher's definition be present.  For example, some types of
mathematical structures generalize others by simply increasing the number of
dimensions.  Since these need not lead to new expressions that both preserve
and relax some of the constraints on the old expression, they need not be a
type of generalization that fits Kitcher's definition, and thus would not be
the
type of structures with the heuristic role indicated by our position.

 \section{Quaternions: Nineteenth Century}

In this section a number of the significant mathematical developments
associated with quaternions during the nineteenth century will first be
considered.  The degree to which they may be seen to have played any sort of
heuristic role in physical developments in that century will then be
considered.

 \subsection{Mathematical developments}

Quaternions were discovered  by Hamilton in 1843 after more than a decade
of attempts to generalize complex numbers to three dimensions.  Instead of
an entity  which he expected to be characterized by three numbers, Hamilton
found that four numbers were required.\footnote{
Hamilton's discovery forms one of the well documented discoveries in
mathematics. Details may be found in biographies of Hamilton by
Hankins~\cite{Hankins;1980} and  O'Donnell~\cite{O'Donnell;1983},
and in the more specialized studies in
Refs.~\cite{Altmann;1986,Altmann;1989,Crowe;1967,Waerden;1976}
{}~\cite{Waerden;1985,Whittaker;1940,Whittaker;1945}.  For Hamilton's
work see his {\it Lectures on Quaternions\/}~\cite{Hamilton;1853},
and the collection of papers pertaining to quaternions in
Ref.~\cite{Hamilton;1967}.}
In  modern notation, Hamilton discovered that the form
\be
 \mbox{\boldmath $q$}  = q_{0} +  q_{1}\mbox{\boldmath $e$}_{1} +
q_{2}\mbox{\boldmath $e$}_{2} + q_{3}\mbox{\boldmath $e$}_{3}
\label{definitionq}
\ee
with multiplication rules for the ``quaternion units''  $\mbox{\boldmath
$e$}_{i}$  given by,
\be
\mbox{\boldmath $e$}_{i}\mbox{\boldmath $e$}_{j} = -\delta_{ij} +
\epsilon_{ijk}\mbox{\boldmath $e$}_{k}
\label{definitione}
\ee
obeys the same multiplication rules as complex numbers except for
commutativity.  In equation~(\ref{definitione})  $\epsilon_{ijk}$ is
antisymmetric in the indices with $\epsilon_{123} = \epsilon_{231} =
\epsilon_{312} = 1$, and the summation convention is assumed for repeated
indices.  With these rules the quaternion product has the form
\be
\mbox{\boldmath $p$} \otimes \mbox{\boldmath $q$}  =  (p_{0} q_{0} -
p_{i} q_{i}) + p_{0}q_{i}\mbox{\boldmath $e$}_{i}  +
p_{0}q_{i}\mbox{\boldmath $e$}_{i}  +
\epsilon_{ijk}p_{i}q_{j}\mbox{\boldmath $e$}_{k}.
\label{product}
\ee
 Hamilton defined a conjugate quaternion by
\be
 \mbox{\boldmath $\overline{q}$} = q_{0} -  q_{1}\mbox{\boldmath $e$}_{1}
- q_{2}\mbox{\boldmath $e$}_{2} - q_{3}\mbox{\boldmath $e$}_{3}
\label{definitioncq}.
\ee
With the norm of a quaternion given by
\be
N(\mbox{\boldmath $q$}) \equiv \mbox{\boldmath $q$} \otimes
\mbox{\boldmath $\overline{q}$}  = q_{0}^{2} +  q_{1}^{2}  + q_{2}^{2} +
q_{3}^{2},
\label{definitonnorm}
\ee
 an important ``law of moduli'' holds for two quaternions:
\be
 N(\mbox{\boldmath $p$} \otimes \mbox{\boldmath $q$})   =
N(\mbox{\boldmath $p$})N(\mbox{\boldmath $q$}).
\label{lawmodulii}
\ee
The law of moduli was of particular importance to Hamilton and it occurs in
a notebook entry made on the very day he discovered quaternions.  Indeed in
a letter written the day after his discovery he noted that without this
property
he would have considered the ``whole speculation as a failure.''  With this
property quaternions can be divided and form a division algebra.    Scalar and
vector parts of a quaternion can be defined as
\be
 S\mbox{\boldmath $q$} = \frac{1}{2}(\mbox{\boldmath $q$} +
\mbox{\boldmath $\overline{q}$}), \hspace{.7cm}
V\mbox{\boldmath $q$} = \frac{1}{2}(\mbox{\boldmath $q$} -
\mbox{\boldmath $\overline{q}$}),
\label{defvs}
\ee
  and it may be readily seen that for a product of two ``pure quaternions''
consisting only of vector parts, the multiplication law in
equation~(\ref{product}) contains in one product both  the ``dot'' and
``cross''
products of the vector analysis that was later to be developed.  From the
beginning Hamilton was concerned with the geometrical interpretation of
quaternions and sought a role for  quaternions in describing rotations of
pure quaternions.  Cayley~\cite{Cayley;1845}, however, was the first to
publish what is now the accepted understanding. If $\mbox{\boldmath $R$}$
is a quaternion of norm 1, then a rotation of $\mbox{\boldmath $q$}$ given
by
\be
  \mbox{\boldmath $q'$} = \mbox{\boldmath $R$}\mbox{\boldmath
$q$}\mbox{\boldmath $R$}^{-1},
\label{rotation}
\ee
 leaves  S$\mbox{\boldmath $q$}$ invariant and transforms
V$\mbox{\boldmath $q$}$ according to a rotation about an axis given by
the pure vector part of $\mbox{\boldmath $R$}$.  In particular if
$\mbox{\boldmath $R$}$ is parametrized by
\be
\mbox{\boldmath $R$} =  \cos\frac{\alpha}{2} + (r_{1}\mbox{\boldmath
$e$}_{1} + r_{2}\mbox{\boldmath $e$}_{2} + r_{3}\mbox{\boldmath
$e$}_{3}) \sin\frac{\alpha}{2}
\label{paramrot}
\ee
then the rotation consists of a rotation of $\alpha$ about  an axis
determined by the direction of  $(r_{1}\mbox{\boldmath $e$}_{1} +
r_{2}\mbox{\boldmath $e$}_{2} + r_{3}\mbox{\boldmath $e$}_{3})$.   Cayley
noted that the parameterization of the transformation in
equation~(\ref{paramrot}) corresponded to that given by Rodrigues in 1840
three years before Hamilton's discovery of quaternions.\footnote{As
Altmann's work~\cite{Altmann;1986,Altmann;1989} has clearly revealed,
Hamilton mistakenly gave a preference to interpreting quaternions in the
form of equation~(\ref{paramrot}) as representing rotations by $\alpha/2$
rather than by $\alpha$.   Associated with this interpretation were two
further problems of interpretation.  First, Hamilton associated the quaternion
units, $\mbox{\boldmath $e$}$, with $\pi/2$ rotations following the
interpretation of the complex $i$ representing such a rotation in a
2-dimensional Argand plane.  Instead the units should be associated with a
rotation by $\pi$.  Second, he identified  a pure quaternion as  a vector in
3-dimensional space instead of what we now know to be its proper
identification  as  a pseudovector.}  Cayley also showed that rotations in a
4-dimensional Euclidean space could be given by quaternions.

In his early writings on quaternions Hamilton also introduced the
3-dimensional ``del'' operation which he wrote as $\lhd$:
\be
\lhd = \mbox{\boldmath $e$}_{1}\frac{d}{dx}  +  \mbox{\boldmath
$e$}_{2}\frac{d}{dy} +  \mbox{\boldmath $e$}_{3}\frac{d}{dz},
\label{del}
\ee
and noted of its square,
\be
- {\lhd}^{2} = {\left(\frac{d}{dx}\right)}^{2} +
{\left(\frac{d}{dy}\right)}^{2}  +
{\left(\frac{d}{dz}\right)}^{2}
\label{del2}
\ee
that ``applications to analytical physics must be extensive to a high
degree.''\footnote{For further details see Ref.~\cite{Hamilton;1967}, p.\
263.}
Until his death in 1865 Hamilton devoted most of his work to developing the
mathematical properties of quaternions without, however, considering much
in the way of their applications to physics.

Shortly after Hamilton's discovery of quaternions a generalization to eight
units was discovered by Graves and independently by Cayley.  These
``Octonions'' obeyed the ``law of moduli'' of equation~(\ref{lawmodulii}) but
without the
algebraic property of an associative multiplication law.   Before the century
had closed two important results clarified further the nature of
generalizations involved in quaternions and octonions.  In 1878
Frobenius~\cite{Frobenius;1878}  proved that the only associative division
algebras consist of the real, complex, and quaternion numbers, and in 1898
Hurwitz~\cite{Hurwitz;1898} proved that if associativity is dropped only one
further division algebra results, viz., that of the octonions.  A further
generalization
took place when quaternions were shown to be a particular example of order
three of Clifford algebras~\cite{Waerden;1985}; however, higher orders of
Clifford algebras fail to form a division algebra which is the property
quaternions share with the real and complex numbers.

In the light of the principle we are proposing here it is interesting to note
that
for a number of  nineteenth century partisans of quaternions that followed
Hamilton, such as the Edinburgh physicists Tait and Knott, quaternions were
very much seen to be generalizations of previous mathematical structures.
Kelland and Tait~\cite{Kelland;1882}, for example, noted explicitly that
quaternions provided ``the most beautiful example of extension by the
removal of limitations'' and noted how room was made for a new
understanding of multiplication once the commutativity law was given up by
Hamilton.  Both of these features  correspond to elements of the
characterization of generalization given by Kitcher.   Furthermore, Kelland
and Tait saw
these features as reasons to consider the application of quaternions to
physics.
We
note at this point one
important aspect of Kitcher's definition of mathematical generalization that
is exemplified by quaternions, namely,  since such generalizations  both
preserve and relax a number of the features of the previous structure they
will entail a limited set of new features.  It is the limited options available
with such particular structures which provides structures of interest for
forging
new physical theories.  We will return to this point later.

 \subsection{Physical Applications}

The main use of quaternions in the nineteenth century consisted in
expressing physical theories in the notation of quaternions rather than in
Cartesian coordinates.  One of the important works where this was done was
Maxwell's {\it Treatise on Electricity and Magnetism\/}~\cite{Maxwell;1891}.
As
well as presenting equations in Cartesian coordinates in a number of places
Maxwell also gave their quaternionic form.   Of particular importance was his
use of Hamilton's ``del'' operator of equation~(\ref{del}).  We find no
examples where they played a role in the development of new physical
theories.  While one does find claims, such as those in a textbook by
McAulay~\cite{McAulay;1893}, that new results from existing theories were
obtained by the use of quaternions they were of a relatively minor nature.
The importance of the notational role of quaternions, however, should not be
underestimated.  Again and again those who used quaternions such as Tait
and McAulay  emphasized that the physical meaning of equations was
revealed in a transparent manner when they were expressed in quaternionic
form.\footnote{
For Tait's comments on this aspect of quaternions see Chapters CXVI
and CXVII of his collected works~\cite{Tait;1831} and
Ref.~\cite{Tait;1890}.}
One finds echoes of this virtue ascribed to quaternions in the nineteenth
century also in recent years when mention is made of the value of using
coordinate free methods of modern differential geometry in spacetime
physics, rather than the older tensor methods.

In addition, the quaternionic formulation, and especially as used by Maxwell
in his {\it Treatise},  did play an important role in the independent
development of the vector analysis by Gibbs and Heaviside in the 1880's.
Both Gibbs and Heaviside noted that a formulation of electromagnetism
could be given using the separate vector and scalar parts of Hamilton's
quaternions, and moreover, such a formulation proved to be far more
accessible for the individual representation of electric and magnetic effects.
Heaviside~\cite{Heaviside;1950}, for one, emphasized what he felt to be the
impractical nature of quaternions and when referring to the negative norm
for a vector, when taken as part of a quaternion, wrote of the ``inscrutable
negativity of the square of a vector in quaternions; here, again, is the root
of
the evil''~\cite{Heaviside;1893}.  The analyses of
Stephenson~\cite{Stephenson;1966}, Crowe~\cite{Crowe;1967}, and
Hankins~\cite{Hankins;1980} provide details of the important role
quaternions played in the later developments of vector analysis.      Towards
the end of the century the very value of using quaternions at all in physics
gave rise to an interesting and rather heated series of exchanges in the
journal
{\it Nature} between Tait, McAulay, and Knott on the one side, as supporters
of quaternions, and Gibbs and Heaviside on the other.  A recent biographer of
Heaviside~\cite{Nahin;1987} has referred to this debate as the ``The Great
Quaternionic War''.\footnote{
Details of this debate and references to the original literature may be
found in Refs.~\cite{Bork;1966,Nahin;1987}}
At the end of the century the methods of vector analysis had become
standard, and the value of quaternions largely discredited.

Writing in 1943 on the occasion of the centenary of the discovery of
quaternions, Whittaker
noted that one of the reasons for the demise of the ``Hamilton school''
consisting of those that sought to make quaternions central in physics, was
their failure to continue ``that instinct for the generalization of a theory
which is characteristic of the mathematician''~\cite{Whittaker;1940}.
Whittaker suggested that people such as Tait and Knott focused more on the
``re-writing'' of existing theories than on exploiting the significant and
unique
aspects of quaternions such as their non-commutativity properties.  In
addition, Whittaker noted, formal mathematical developments related to
physical ideas which to some physicists represent ``mere mathematical
playthings'' tend
later to come into prominence in physics.  One example Whittaker
mentioned was the way many of the more mathematical aspects of
Hamilton's work on dynamics have found a place in quantum theory in this
century.    Whittaker's comments underline our position in attesting to an
aspect of the particular heuristic role we are assigning to mathematics in
physics.

 \section{Quaternions: Twentieth Century}

It has only been in this century that unique features of quaternionic
structures
have been woven closely into the development of new physical theories.
Two particular examples which we mention in Section 4.2 are the theories of
quaternionic quantum mechanics and  supergravity.  A number of uses, such
as some of the applications to special relativity, do clearly fit the category
of
``reformulation'' of existing theories with claims that the resultant structure
possess an elegance and aesthetic appeal.

 \subsection{Mathematical Developments}

Three particular mathematical developments associated with quaternions
will be mentioned here.  All three associate quaternions with other important
mathematical structures, and represent a phenomenon in the development
of mathematics which the historian of mathematics
Wilder~\cite{Wilder;1968,Wilder;1981} has referred to as ``consolidation.''
When consolidation takes place various structures which were originally
separate from
each other, and seemingly unrelated, are brought into relationship with each
other.   The importance for physics resides in presenting various ways in
which a generalized mathematical structure may be seen to be related to those
within existing theories, and thus may be used in new theories.

 The first is the association of quaternions with the Pauli matrices which is
mathematically rather insignificant, but of some importance for physics.  In
the nineteenth century   Cayley had given a matrix representation of
quaternions, but one important   realization of the  quaternionic units,
$\mbox{\boldmath $e$}$, is that given by $\mbox{\boldmath $e$}_{i} = -
i\mbox{\boldmath $\sigma$}_{i}$ where $\mbox{\boldmath
$\sigma$}_{i}$ are the 2 x 2 Pauli matrices.   Pauli noted in his paper of
1927,
where he introduced the matrices,  that Jordan had pointed this out to him.
Thus a quaternion can be represented by,

\be
 \mbox{\boldmath $q$}  = q_{0} -  iq_{i}\mbox{\boldmath $\sigma$}_{i}
= \left[ \begin{array}{cc}
a & - b^{\star}\\
b & a^{\star}
\end{array} \right],
\label{Pauli1}
\ee
where,
\be
a = q_{0} - iq_{3}, \hspace{.5cm} b = q_{2} - iq_{1}.
\label{Pauli2}
\ee
In equation~(\ref{Pauli1}) the $^{\star}$ represents complex conjugation.
Thus a neat correspondence is obtained between quaternions and the SU(2)
Lie group.  In addition the operator $U(\alpha)$ that rotates the
2-dimensional spinor representations of the rotation group by $\alpha$ about
a
direction $\hat {\mbox{\boldmath $n$}}$  is given by~\cite{Sachs;1982}:

\be
 U(\alpha)  =  \exp(i\frac{\mbox{\boldmath $\sigma$} \cdot   \hat
{\mbox{\boldmath $n$}}}{2})
 = \cos (\frac{\alpha}{2})  -  i \mbox{\boldmath $\sigma$} \cdot   \hat
{\mbox{\boldmath $n$}} \sin (\frac{\alpha}{2})
\label{rotdem}
\ee
 The similarity of this transformation with the Cayley-Rodrigues
parametrization in equation~(\ref{rotation}) is immediately evident.  In
addition, equation~(\ref{rotdem}) may readily be seen as a generalization of
de Moivre's theorem for complex numbers.

The second development we wish to mention is the association between
division algebras and Lie groups.   Through the work of Freudenthal,
Rozenfeld and Tits, which may be found summarized in a review by
Freudenthal~\cite{Freudenthal;1965}, and in an application to supergravity
by G\"unaydin {\it et al.\/}~\cite{Gunaydin;1983}, it was realized that the
four categories of
semi-simple Lie groups, viz.,  orthogonal,  unitary,  symplectic and
exceptional groups,  were associated with the real numbers, complex
numbers,
quaternions and octonions via what is known as the ``magic square.''
Commenting on this rather remarkable result, G\"ursey notes, that it
associates the division algebras with ``the very core of the classification of
possible symmetries in nature''\cite{Gursey;1983}.

The third result is a beautiful relationship between the introduction of
coordinates in affine and projective planes  and the complex, quaternion, and
octonion numbers~\cite{Artmann;1988}.   With an affine plane associated
with  complex numbers the projective theorem of Pappus holds,
whereas it does not for an affine plane associated with quaternionic
structures.  And while the theorem of Desargues holds for the latter plane,
for
the affine plane associated with octonionic structures neither the theorem of
Desargues nor that of Pappus holds, although a more restricted form of the
theorem of Desargues may be proved.   This more restricted form was proved
by Ruth
Moufang in 1933.

\subsection{Physical Applications}

\subsubsection{Quaternions and Spacetime Physics}

The presence of four units in quaternions posed a problem of interpretation
to Hamilton.   Initially he had vaguely surmised that the vector part of the
quaternion could be likened to a sort of ``polarized intensity'' while the
scalar
part to an ``unpolarized energy.''  Then in a letter to a friend in 1844 he
wondered whether the vector part could represent the three space
dimensions and the scalar part represents time.\footnote{References to
these
passages may be found in Hankin's biography~\cite{Hankins;1980}, p.\ 301.}
His latter view has been the way quaternions have been used to formulate
special relativity, and their mathematical properties allow elegant expressions
to be derived for all the expressions in special relativity.

 Quaternions were first introduced into special relativity by Conway in
1911~\cite{Conway;1911} and independently a year later by
Silberstein~\cite{Silberstein;1912}.  There has been a long tradition of using
quaternions for special relativity and a review by Synge~\cite{Synge;1972}
covers developments up until the 1960's.  Modern presentations have been
given by  Edmonds~\cite{Edmonds;1972,Edmonds;1974},
Sachs~\cite{Sachs;1982}, Gough~\cite{Gough;1989} and Abonyi {\it et
al.,\/}~\cite{Abonyi;1991}.  The use of quaternions in special relativity,
however, is not entirely straigtforward.  Since the field of quaternions is a
4-dimensional Euclidean space, complex components for the quaternions are
required for the 3 + 1 spacetime of special relativity.  Quaternions of this
nature were called biquaternions by Hamilton,  and do not form a division
algebra.  Also there is a choice as to whether to express the scalar or the
vector
part of the quaternion in complex form.  With the latter convention a
spacetime point, $(ct, x_{1}, x_{2}, x_{3})$, can be expressed as the
quaternion
\be
\mbox{\boldmath $x$}   =  ct + i x_{i}\mbox{\boldmath $e$}_{i}.
\label{stq}
\ee
A Lorentz transformation of a boost, for example, of $v$ in the $x_{1}$
direction can be written as~\cite{Silberstein;1912,Synge;1972,Gough;1989}
\be
\mbox{\boldmath $x$}' = \exp (i\frac{\mbox{\boldmath $e$}_{1} \theta}{2})
\mbox{\boldmath $x$} \exp(- i\frac{\mbox{\boldmath $e$}_{1} \theta}{2}),
\hspace{.7cm} \tanh \theta = \frac{v}{c}.
\label{LT}
\ee
Such a transformation leaves the norm $\mbox{\boldmath $x$} \otimes
\mbox{\boldmath $\overline{x}$} = (ct)^{2}  - x_{i}x_{i}$ invariant. Similar
expressions may be formed for other spacetime quantities such as a four
momentum and electromagnetic potentials, and electrodynamics can readily
be given a quaternionic formulation~\cite{Majernik;1976,Abonyi;1991}.

While an elegant reformulation of the equations of special relativity alone
provides a motivation for the use of quaternions, various related avenues of
research have emerged from this context.   Rastall~\cite{Rastall;1964},
Singh~\cite{Singh;1982}, and Sachs~\cite{Sachs;1982}, for example,  have
shown there are certain advantages in representing field equations such as
the Dirac equation and Maxwell equations in quaternionic form when a
generalization is made to Riemannian space-time.

 Various quaternionic formulations of Dirac's relativistic equation have been
considered  the 1930's onwards.  Early presentations may be found in
Conway's work~\cite{Conway;1937}, and more recent presentations in the
work of Edmonds~\cite{Edmonds;1974}, Gough~\cite{Gough;1989}, and
Davies~\cite{Davies;1990}.   When written in this manner a doubling of the
components of the wavefunction from four to eight occurs and the possible
physical significance of these components has been a matter of
speculation~\cite{Edmonds;1974,Gough;1989}.
 Adler~\cite{Adler;1989b}, for example, has exploited this feature to develop a
novel form of QED which eliminates the need for a Dirac sea of negative
energy electrons by combining both particle and antiparticle states within a
single species of fermion.  In addition, in an interesting paper
Davies~\cite{Davies;1990} has shown that when potentials  are included in a
quaternionic Dirac equation certain restrictions on their components are
required which raise questions as to the observability of effects unique to a
nonrelativistic quaternionic quantum mechanics.

 Finally we note that the quaternionic formulation of spacetime theories has
been extended to superluminal Lorentz transformations.
Imaeda~\cite{Imaeda;1976,Imaeda;1979} has presented such a transformation
for a boost in the $x_{1}$ direction in the form
\be
\mbox{\boldmath $x$}' = \pm i \exp (i\frac{\mbox{\boldmath $e$}_{1}
\theta}{2}) \mbox{\boldmath $x$}  \exp (- i\frac{\mbox{\boldmath $e$}_{1}
\theta}{2})  \mbox{\boldmath $e$}_{1}, \hspace{.7cm} \coth\theta =
\frac{v}{c},  \hspace{.7cm} (\frac{v}{c})^{2} >1
\label{SLT}
\ee
Recently Teti has given an expression unifying both the subluminal and
superluminal Lorentz transformations~\cite{Teli;1980}.

In these examples we can see how  the extra structure of quaternions over
complex and real numbers has enabled new perspectives through permitting
different formalisms, and moreover, provided structures within which new
physical theories can be considered.

\subsubsection{Quaternionic Quantum Mechanics}

In an important paper in 1936 Birkhoff and von
Neumann~\cite{Birkhoff;1936} presented a propositional calculus for
quantum mechanics, and noted that  a concrete realization leads to a general
result that a quantum mechanical system may be represented as a vector space
over the real, complex and quaternionic fields.    Their paper was the first to
point out the possibility of a quaternionic formulation of  quantum
mechanics.   Their result essentially means that the quantum mechanical
superposition principle for probability amplitudes only determines the
quantum mechanical probabilities to obey the ``law of moduli'' and thus to
be one of the division algebras, and not necessary to be the algebra of complex
numbers~\cite{Adler;1992}.  With a quaternionic extension the wavefunction
may be given the form
\be
\Psi = \Psi_{0} + \Psi_{i} \mbox{\boldmath $e$}_{i},
\label{acrazywf}
\ee
where $\Psi_{0}$ and $\Psi_{i}$ are real.  One can proceed to develop a
quaternionic quantum mechanics (QQM) with states defined on a
quaternionic Hilbert space.   With such an extension  the rays in the Hilbert
space representing pure states are no longer  one dimensional subspaces and
the c-numbers no longer commute.

Studies  on the application of quaternions in quantum theory go back to the
1950's.  The possibility of using quaternions as a basis for a field theory was
considered by C. N. Yang in 1957.   Since the phase in the complex algebra is
associated with electromagnetism, Yang's idea was to see if a phase in
quaternionic algebra could be related to isotopic spin gauge fields that might
then account for the existence of isotopic spin symmetry.   While this hope
was not fulfilled  Yang noted in 1983 that he still believed the direction was
a
correct one.\footnote{Yang's account of these attempts may be found in the
introductory commentary to a collection of his papers~\cite{Yang;1983}. p. 22-
-23.}     In the late 1950's and early 1960's several aspects of QQM were
investigated in a series of foundational papers by Finkelstein {\it et al.\/}
\cite{Finkelstein;1962,Finkelstein;1963,Finkelstein;1963a,Finkelstein;1979}.
Contemporary presentations of QQM  may be found in the works  of
Adler~\cite{Adler;1980,Adler;1986,Adler;1988,Adler;1989b,Adler;1992},
Horwitz and Biedenharn~\cite{Horwitz;1984}, and Nash and
Joshi~\cite{Nash;1987,Nash;1987a,Nash;1988,Nash;1992}.

 QQM provides an excellent illustration of our principle as to how
generalizations of certain mathematical structures can provide the avenues to
explore new physical theories.   QQM has many features which make it a far
richer theory than complex quantum mechanics.  It is not simply a matter of
increasing the internal degree of freedom of one of the variables in the
conventional complex theory, and defining many of the notions that
correspond to the conventional theory has proved to be a difficult and
interesting task.   In particular, we mention the following issues which arise
as  unique concerns for QQM.  To begin with, the proper generalization of the
Schrodinger equation  was by no means clear.
  Simply replacing  the
imaginary $i$ of complex quantum theory in the Schr\"odinger equation
with $\mbox{\boldmath $e$}$ proved not to provide the proper
generalization.  Rather, it was found the proper form for the Schr\"odinger
equation is given by~\cite{Adler;1986,Davies;1989},
\be
\frac{d\Psi}{dt} = - \overline{H} \Psi
\label{acrazySE}
\ee
where  $\overline{H}$ is a quaternion---anti-self-adjoint Hamiltonian.  In
addition, there is a problem defining the tensor product of wavefunctions for
composite systems due to the non-commutativity of wavefunctions such as
in
equation~(\ref{acrazywf}).   A number of people have taken this as a
reason to rule out  QQM; however, a recent definition by Nash and
Joshi~\cite{Nash;1987} has provided a way to define such products, and
moreover, to define them in a way that suggests  how the effects of  QQM may
be hidden.  Also, there has been the question as to whether  phase
transformations in the case of a quaternionic quantum field theory should be
defined as having the form  $ \phi \rightarrow p\phi$, where $p$ is any unit
quaternion, as Adler~\cite{Adler;1988} indicates, or as  $ \phi
\rightarrow p\phi p^{-1}$ as in the original papers of Finkelstein {\it et
al.,}.
One recent study indicates there may be reasons for preferring the latter
definition~\cite{Nash;1988}.

The relationship of  QQM to complex quantum mechanics remains an
interesting and unresolved issue.  Could QQM apply to the realm of high
energies, for example, or provide a theory for understanding preon dynamics
while complex quantum mechanics only applies to presently observed
particles?  Or could QQM be a ``cover'' theory which applies to all particles,
and reduce in some way to conventional quantum mechanics in realms in
where we have confirmation of the conventional theory?  Related to these
issues is the result of some interest that the correspondence principle of QQM
does not entail a limit to some form of quaternionic classical mechanics but
rather to a form of conventional quantum theory~\cite{Adler;1988}.

	Finally, we should mention that one very significant result of the study
of
QQM was a formulation of a new form of gauge invariance by Finkelstein {\it
et al.,} which they
have labeled as ``Q-covariance''~\cite{Finkelstein;1963a}.   Given a
quaternionic phase transformation of  the form   $ \phi \rightarrow p\phi
p^{-1}$, a statement that all of the physical laws are invariant to such a
transformation leads to the introduction of a set of massive gauge bosons.
There may be a connection between this manner of mass  generation
 and the Higgs mechanism.   If this proves to be true we may obtain new
insights into the profound problem of mass generation.

To achieve a clear resolution of some of these unsolved problems has both
the potential to provide new physical theories, and at the same time to enable
a better understanding of our present conventional quantum mechanics.

 \subsubsection{Some Recent Applications in
Theories of Gravity}

Complex numbers have played an important role in formulating various
theories of gravitation.  Spinors, for example, which are ordered pairs of
complex numbers, provide powerful tools in exploring the structure of
general relativity theory.
  In recent years various ways of extending the geometrical structure of
general
relativity have been considered such as in supergravity and Kaluza-Klein
theories.   Quaternions have also been used as a way to generalize the
geometrical structure.   For example, in standard general relativity theory the
metric is a real bilinear form on a tangent space at each point in the
spacetime
manifold.   Various extensions of the tangent bundle to other spaces have
been considered such as to complex numbers and hypercomplex numbers.
Mann~\cite{Mann;1984} has recently considered a theory in which the
tangent bundle is extended from a field of real numbers to one of
quaternions.
Mann's approach introduces a
non-Abelian framework into spacetime structure, but has the advantage of
leaving
features of conventional general relativity unchanged.  In particular the
spacetime manifold is real as well as quantities such as the invariant interval
$ds^{2}$ defined on it.

Quaternionic structures have also been recently used to provide a possible
framework in which to consider quantum theories of gravity.   The work of
Witten~\cite{Witten;1988,Witten;1988a,Witten;1988b}, for example, has
shown the
potential importance of topological considerations for the exploration of
quantum field theories of relevance for general relativity.    In addition,
various elegant studies of
3-dimensional formulations of
gravity as a gauge theory using a Chern-Simons action have been
given~\cite{Deser;1982,Achucarro;1986,Horne;1989,Witten;1988}.  The
geometrical structures provided by quaternions may allow various
4-dimensional formulations to be obtainted.  Certainly  quaternions  have
surfaced in considerations of $N=2$ supergravity, as the geometric structure
of quaternionic manifolds is of interest to theories such as the
non-linear sigma models that appear within supersymmetric
theories~\cite{deWit;1991a,deWit;1991}.

 The theoretical and experimental consequences of these studies are uncertain
at the moment, however, we see how both the generality of the
structures provided by quaternions, as well as the particular dimensions of
their algebra are providing  those structures of
interest to the four dimensionality of spacetime.

  \subsubsection{Quaternions in Applied Physics}

It is worth drawing attention to the rather remarkable way in which
quaternions have
emerged in recent decades in several  applied areas outside of theoretical
physics.   While strictly outside the theme of this essay, we wish to mention
some examples as their use in  such areas will undoubtedly have an influence
on their place within the physics of the future.  It is often due to the use of
certain mathematical structures in technological situations that they become
part of the textbook tradition and the teaching of basic disciplines.  One area
of
application arises  from their excellent ability to represent rotations in
three
dimensional space, and the other through certain analytic properties of
functions defined over quaternions.   The contemporary practical applications
of quaternions would have surprized some of the nineteenth century
adversaries of quaternions.  Heaviside in particular had noted that ``it is
practically certain that there is no chance whatever for Quaternions as a
practical system of mathematics \ldots''~\cite{Heaviside;1893}, and Cayley
stated that they seem ``a very artificial method for treating such parts of the
science
of three-dimensional geometry''\cite{Cayley;1894}.

In particular, the recent studies by Tweed {\it et
al.\/}~\cite{Tweed;1987,Tweed;1990,Tweed;1990a}, building on pioneering
work of Westheimer~\cite{Westheimer;1957} in 1957, have used
quaternionic algebra to represent the intricate rotational motion in eyes
movements.  Both Westheimer and Tweed {\it et al.\/} note the advantages
of computational efficiency as well as simplicity of expression when
quaternions are used in the formulation of laws governing eye movements
such as Listing's Law.   Quaternions have also been used for calculations
needed in robotic
control~\cite{Walker;1991,Young;1990,Chou;1991,Chou;1992}, computer
graphics~\cite{Seidel;1990}, and in determining spacecraft
orientation~\cite{Wertz;1980}.  The spaceshuttle's flight software, for
example, uses quaternions in its computation for guidance navigation and
flight control.   The advantages of the parameterization by quaternions
(usually referred to as the ``Euler parameters'')  over other means such as the
Euler angles include: i) speed of calculation; ii) avoidance of singularities;
iii)
providing a minimum set of parameters;  iv) enabling other physical
quantities such as angular momentum to be derived from the quaternionic
parameterization in a particularly simple manner.

 A second category of  applications draws on the analytic properties of
functions
of quaternions that were investigated by Fueter in the
1930's~\cite{Fueter;1935,Fueter;1936,Fueter;1937}.  A presentation of Fueter's
work in English may be found in a study by Deavours~\cite{Deavours;1973}.
Many of the results of complex analysis such as the Cauchy and Liouville
Theorems generalize to quaternionic analysis and the powerful
 two dimensional results of complex analysis can be extended to
 three dimensions.  In particular, this has recently been applied in a study
by
Davies {\it et al.,\/}~\cite{Davies;1989a} to the derivation of integral
transforms of vector functions in
 three dimensions with an illustration of geophysical interest.  It appears
that
quaternions form a natural
co-ordinate system for vector integral transforms in
 three  dimensions, and it may be surmised that using this feature of
quaternions will provide a profitable approach to the study of integral
transforms~\cite{Zhdanov;1987}.   The results are   of  immediate interest in
areas such geophysical exploration and remote sensing where integral
transforms of fields in three dimensions are used.

 \section{Conclusions}

Our principle of the heuristic role of mathematical structures in physics
specifies
that those structures which are generalizations of structures currently part of
successful physical theories will be  the ones  well suited for the
development
of new physical theories.
 Quaternions provide an illustration of some complexity, and in addition the
status of the contemporary theories of which they are a part of is uncertain at
the moment.  Nevertheless several features emerge from this illustration
that we would surmise occur whenever mathematical structures of this sort
are used in physics.

First, the use of the generalized formalism provides structures from which
{\it new mathematical formalisms\/} of use for physical theories may
emerge. The manner in which vector analysis, with its separate cross and dot
product, arose from the quaternionic product is one example of this
occurring.

Second, often elegant ways of stating familiar results occur.  The
use of quaternions in any situations involving rotations in three or four
dimensions, such as relativity theory expressed in Euclidean space, provide
an example of how this may happen. In these cases there is cause to claim that
the physical situation is revealed in a particularly clear manner by the
formalism.

Third, attempts to use the formalism in new physical theories has potential to
reveal more about the experiment and theoretical status of the theories which
use the mathematical structures from which the generalization occurred.
There are reasons to be optimistic that QQM will play this role.  Even its
failure to provide any viable physical theory has potential to illuminate the
vital role complex numbers play in the theory.

Fourth, and indeed this is a rather important point, new physical theories can
emerge using the generalized structures in a way that preserves many of the
virtues of the theories associated with the previous mathematical structure.
Again QQM may play a role such as this, and the various ways in which
quaternions
seem to be appearing in supergravity theories indicate a role for them of this
nature.

Finally it is important to note that the particularity of the generalizations
we
are concerned with is one of the reasons they are productive in the search for
new physics.  The balance between each of Kitcher's conditions for
generalization ensures such a property.  The generalized structures provide
particular relationships which capture the particular features of nature.
Mathematics plays an important role by limiting the possibilities for the
physicist
to consider.\footnote{Zee, for example, stresses the value of this feature of
mathematics, and notes that the limited number of Lie algebras is of
extraordinary help in constructing GUTs~\cite{Zee;1990},
p.\ 312.}   Quaternions generalize other mathematical
structures in Kitcher's sense, but  have a particular
 four dimensional structure.  Physicists who have used quaternions have
noted this point.  Tait, for example,  noted in 1894, in response to an attack
on
quaternions due to their limited number of dimensions, that from a physical
point of view this is not a defect, but ``is to be regarded as the greatest
possible
recommendation.''  For Tait it showed them to be particularly relevant to the
``actual'' world~\cite{Tait;1894}.   And 70 years later Rastall remarked that
contrary to the spirit of certain rarefied mathematical approaches to spacetime
theories, quaternions are useful to those ``prepared to exploit the accident of
having being born in
space-time''~\cite{Rastall;1964}.

 There is good reason then to see the active exploitation of certain
mathematical generalizations as providing good guides for the physics of the
future.

\newpage



\end{document}